\documentclass[runningheads]{llncs}
\usepackage{graphicx}
\usepackage{subfiles}

\usepackage{todonotes}
\usepackage{booktabs}
\usepackage{multirow}
\PassOptionsToPackage{hyphens}{url}\usepackage{hyperref}

\usepackage{dsfont}
\usepackage{amsmath}

\usepackage{caption}
\usepackage{subcaption}

\usepackage[numbers,sort&compress]{natbib}

\usepackage{xcolor}

\begin{document}
\title{Goldilocks: Just-Right Tuning of BERT for Technology-Assisted Review}
\author{
Eugene Yang\inst{1} \and 
Sean MacAvaney\inst{2} \and 
David D. Lewis\inst{3} \and 
Ophir Frieder\inst{4}
}
\authorrunning{E. Yang et al.}
\institute{
HLTCOE, Johns Hopkins University, USA ~\email{eugene.yang@jhu.edu}
\and
University of Glasgow, United Kingdom~\email{Sean.MacAvaney@glasgow.ac.uk}
\and
Reveal-Brainspace, USA~\email{ecir2022paper@davelewis.com}
\and
IRLab, Georgetown University, USA~\email{ophir@ir.cs.georgetown.edu}
}
\maketitle              %
\begin{abstract}
Technology-assisted review (TAR) refers to iterative active learning workflows for document review in high recall retrieval (HRR) tasks. TAR research and most commercial TAR software have applied linear models such as logistic regression to lexical features.  
Transformer-based models with supervised tuning are known to improve effectiveness on many text classification tasks, suggesting their use in TAR. We indeed find that the pre-trained BERT model reduces review cost by 10\% to 15\% in TAR workflows simulated on the RCV1-v2 newswire collection. In contrast, we likewise determined that linear models outperform BERT for simulated legal discovery topics on the Jeb Bush e-mail collection. This suggests the match between transformer pre-training corpora and the task domain is of greater significance than generally appreciated. Additionally, we show that \textit{just-right} language model fine-tuning on the task collection before starting active learning is critical. Too little or too much fine-tuning hinders performance, worse than that of linear models, even for a favorable corpus such as RCV1-v2.

\end{abstract}

\section{Introduction}

\textit{High recall retrieval} (HRR) tasks (also called \textit{annotation} tasks) involve identifying most or all documents of interest in a large collection. HRR tasks include electronic discovery in the law (eDiscovery)~\cite{baron2016perspectives}, systematic review in medicine~\cite{wallace2010semi, clef2017ehealth-tar, clef2018ehealth-tar, clef2019ehealth-tar}, document sensitivity review~\cite{mcdonald2018active}, online content moderation~\cite{content-moderation-self}, and corpus annotation to support research and development~\cite{zhu2010confidence}.

\textit{Technology-assisted review} (TAR) refers to the automated methods to reduce the number of documents reviewed in HRR projects~\cite{oard2013information}. Iterative, pool-based active learning of predictive models for review prioritization is the most commonly applied workflow~\cite{cormack2014evaluation,cormack2016engineering}. 
Linear models such as logistic regression and support vector machines (SVMs) applied to lexical and metadata features are the most common supervised learning approaches. Unlike
in classification and adhoc retrieval tasks, the supervised learning model in TAR is typically discarded after use. This is because each legal case or other project has its own retrieval objective, and because of concerns of leaking confidential information.
Therefore, the cost of training the supervised learning model in TAR often cannot be amortized over future data or across tasks.

Pre-trained transformers~\cite{Vaswani2017AttentionIA} such as BERT~\cite{devlin2018bert}, 
GPT-3~\cite{brown2020language}, and T5~\cite{raffel2020exploring}
are 
effective at a variety of natural language processing tasks.
These models
learn linguistic patterns
from very large corpora in an unsupervised fashion (\textit{pre-training}) and can be tuned to language characteristics of a particular task data set (\textit{LM fine-tuning}) \cite{devlin2018bert, gururangan2020don}. They can then be applied to a task on that data set by zero-shot transfer learning~\cite{macavaney2020sledge, wang2019cross} or by \textit{task fine-tuning} to labeled training data~\cite{hou2020few, yang2020context}.
Transformers have improved effectiveness at tasks related to HRR such as document classification~\cite{adhikari2019docbert}, entity extraction~\cite{devlin2018bert}, and adhoc retrieval~\cite{macavaney2019cedr}. This has inspired initial commercial use of transformers by eDiscovery providers, though not yet in an active learning context.\footnote{\url{https://www.nexlp.com/blog/nexbert-story-engine-cloud}}

We are not aware of published studies of transformers in TAR workflows. Several studies have evaluated task fine-tuning using active learning \cite{liu2020ltp,shelmanov2019active}, including for text classification tasks \cite{zhang2019ensemble,dor2020active}. These studies, however, have evaluated generalization to new data using training/test splits. HRR, like relevance feedback in adhoc search \cite{ruthven2003survey}, is a transductive setting: evaluation is on the same task corpus from which the training data is selected by active learning.

The transductive setting makes of less importance a key advantage of transformers over traditional methods: their inclusion of language-wide linguistic regularities that might be present in unseen test data. It has already been demonstrated by~\citet{gururangan2019variational} that BERT is more effective when the target task domain is similar to the ones on which BERT was trained (English language books~\cite{zhu2015aligning} and English Wikipedia).  Active learning also reduces transformer advantage, by reducing the labeling cost to learn corpus-specific vocabulary and regularities. Finally, the short useful life of TAR models means limited opportunity to amortize training cost, raising questions about the large computational cost of task fine-tuning for transformers.

The recent TREC-COVID evaluation provides evidence both in favor and against transformers. A SciBERT-based zero-shot reranker of BM25-based text retrieval topped several of the Round 1 evaluation measures \cite{macavaney2020sledgeArXiv,macavaney2020sledge}. On the other hand, another transformer-based effort (which omitted langauge model fine tuning) struggled \cite{lima2020denmark}, a number of other deep learning efforts had mediocre effectiveness, and classic linear models based on lexical features and trained by active learning were highly competitive (leading on one measure)~\cite{wang2020participation,macavaney2020sledgeArXiv}.
Recently, \citet{ioannidis2021analysis} evaluated BERT and PubMedBERT~\cite{gu2020domain} on CLEF eHealth Technology Assisted Reviews in Empirical Medicine Task~\cite{clef2017ehealth-tar, clef2018ehealth-tar}. Despite the claim, \citet{ioannidis2021analysis} considered a simple ranking and classification setting instead of an iterative task.

Against this context, we provide the first demonstration of fine-tuned transfor-mer-based models in the TAR transductive active learning setting. We use BERT~\cite{devlin2018bert} as a representative transformer. We fine-tune the language model to each of two (unlabeled) task corpora using a masked language modeling objective, kick off each prioritization 
task on that corpus with a single positive example, and do task fine-tuning of BERT on each TAR active learning iteration.

Surprisingly, despite the past success stories of BERT in dramatically advancing the retrieval effectiveness, in our work, we found that it only performs on par with the simple logistic regression model due to the transductivity of HRR. On the contrary, under certain scenarios, the BERT model reduces the total reviewing cost, which is the primary objective of HRR tasks. Given its data-hungry property, this cost reduction is counterintuitive but yet very favorable. 
We highlight our contributions in the following,
\begin{itemize}
\item First, we find that language model fine-tuning to the task corpus before active learning is critical, but also that too much of it can be done. 
\item Second, we find language model fine-tuning is not a cure-all for domain mismatch. Our fine-tuned BERT model beats linear models on a data set (RCV1-v2) similar to the text types on which BERT was trained, but falls short when operating
with very different textual characteristics. 
\item Finally, we provide a running time analysis to demonstrate the computational overhead for applying BERT.  
\end{itemize}

\section{Background}

HRR projects typically balance thoroughness versus cost by setting a recall target that is high, but below 100\%. Targets such as 80\% recall are common in eDiscovery~\cite{totalrecall2015} and are sometimes encoded in legal agreements~\cite{cost-structure-paper}. Systematic review often shoots for 95\% recall (on smaller and more homogeneous collections)~\cite{clef2018ehealth-tar, clef2017ehealth-tar}. Recall is defined as the number of relevant documents found among the reviewed documents, divided by the number of relevant documents in the defined collection of interest (e.g., all emails from a set of employees relevant to a legal case, or all biomedical research papers that have passed a keyword screening).

TAR workflows reduce costs by using iterative active learning to prioritze batches of documents for review. One-phase TAR workflows continue this process until a stopping rule indicates that the reviewed documents have met the recall target~\cite{cormack2014evaluation}. Two-phase workflows have a training phase followed by a classification phase (on the same data set), with review done in both phases~\cite{mcdonald2018active, cost-structure-paper}. Designing stopping rules that determine as early as possible that a recall target has been reached is an active research area~\cite{cormack2016scalability, cormack2016engineering, callaghan2020statistical, li2020stop, saha2015batch-mode, wallace2010semi, quantstop-paper, qbcb-paper}, but we design our evaluation to avoid the selection and the error incurred by the stopping rule based on the prior studies in TAR cost evaluation~\cite{cost-structure-paper}.

Evaluation for HRR emphasizes a recall/cost tradeoff rather than the related recall/precision tradeoff. In eDiscovery, \textit{Depth for recall (DFR@x)} is the proportion of the collection reviewed to hit a recall target $x$.\footnote{\url{https://www.gibsondunn.com/wp-content/uploads/documents/publications/Evans-Metrics-that-Matter-Inside-Counsel-1.2015.pdf}}
Systematic review uses \textit{Work saved over sampling (WSS@x)}, which subtracts DFR@x from the expected cost to hit the recall target by random sampling: $WSS@x = x - DFR@x$~\cite{cohen2006reducing}. Some early HRR studies also use R-Precision (precision at $R$ where $R$ is the number of relevant documents)~\cite{totalrecall2015, totalrecall2016} to capture the effectiveness to lower part of the rank as opposed to Precision at 5 or 10 in adhoc retrieval.

However, these evaluation metrics do not consider the cost of obtaining the labels for training documents. In this study, we adapt the cost evaluation of TAR proposed by \citet{cost-structure-paper} to jointly evaluate the effectiveness and the cost of the retrieval results. The total cost of TAR consists of the cost of reviewing (1) the training documents and (2) the minimum number of unreviewed documents ranked by the current classification model for fulfilling the recall target. This cost evaluation approach allows documents in different classes and phases to cost differently, facilitating a more practical HRR evaluation and emphasizing the cost of training the one-time classification model.

Commercial TAR technology relies on traditional text classification approaches such as logistic regression and support vector machines (SVMs)~\cite{yang2017icail, brown2015peeking}, that have been widely studied in both active learning and transductive contexts~\cite{cormack2014evaluation, mcdonald2018active, cost-structure-paper, qbcb-paper, quantstop-paper}. However, the state of the art in text classification has moved to transformer-based models such as BERT~\cite{devlin2018bert}
whose properties in these contexts are less well-understood. This gap in understanding motivates the current study.

\section{Adapting BERT for TAR}

In this section, we describe the adaption of the BERT model to TAR. On a high level, the BERT language model is fine-tuned on the collection of the retrieval interest. At each active learning iteration, we select a set of documents based on the predictions from the model for human review. The acquired labels are fed to the BERT model to perform classification fine-tuning for learning relevancy.

Since the entire task corpus is available before training in TAR, our first step in applying BERT to TAR was language model fine tuning to that corpus. We used the same unsupervised masked language modeling task originally used to train BERT: randomly masking 15\% of the tokens in each sequence and tuning BERT's parameters to predict the missing tokens~\cite{devlin2018bert}. The key question is how much to move BERT's parameters (encoding the linguistic regularities explicit in a mammoth broad domain corpus) toward the task-specific, but less complete, explicit regularities of the task corpus. Our experiments study this by varying the number of epochs (passes through training set) in language model fine-tuning.      

TAR workflows use an active learning method such as relevance feedback~\cite{rocchio1971relevance} or uncertainty sampling~\cite{lewis1994sequential}, where the model trained by supervised learning on iteration $k-1$ is used to the select the batch of documents to be labeled in iteration $k$. The union of labeled batches for iterations $1...k-1$ is the training set for iteration $k$. One random relevant document was selected at the beginning of the process as the \textit{seed document} to initiate the active learning. 
All labeled documents are used for classification fine-tuning of the BERT model. Documents labeled in earlier iterations are visited more by the model based on this classification fine-tuning process. However, based on our pilot study on only fine-tuning model on the newly labeled documents at each iteration, the results were far worse than on all labeled documents.
We use a cross-entropy loss on the binary class label by adding a dense layer on the \texttt{[CLS]} token on top of BERT. We train for a fixed number of epochs, which previous work on active learning for BERT suggests works as well as choosing epoch number using a validation set~\cite{dor2020active}. 

For simple models, training
can be done to convergence from scratch on each iteration (as we do for logistic regression and SVMs in our experiments).  Classification fine tuning for a transformer is computationally expensive, so we instead use the model trained on iteration $k-1$ as the starting point for optimization on iteration $k$.  While this potentially gives more influence to examples selected on the first iteration, adaptive example selection by active learning reduces this effect.

\section{Experiment Setup}

\subsection{Data Sets}

We simulate TAR reviews on two fully labeled collections widely used in HRR studies~\cite{oard2018jointly, cormack2016engineering, cormack2016scalability, yang2018desires, yang2019icail, yang2019sigir, totalrecall2015, totalrecall2016}: RCV1-v2~\cite{rcv1} and the Jeb Bush emails~\cite{totalrecall2015, totalrecall2016}. 

RCV1-v2 consists of 804,414 news stories with coding for 658 economic news categories. 
We use the 45 categories subset established by previous high recall retrieval study~\cite{cost-structure-paper} that spans across three prevalence and three difficulty bins. 
Text from the title and body was concatenated and tokenized using WordPiece. Documents are truncated with 512 WordPiece tokens as the leading passages of the news documents usually convey the most important aspects of the news articles~\cite{catena2019enhanced}. 
The collection is also downsampled to 20\% (160,833 documents) for computational efficiency.

The Jeb Bush collection consists of 274,124 unique emails between the former governor of Florida and his colleagues and constituents. The collection was annotated for 44 political topics for the 2015 and 2016 TREC Total Recall Tracks~\cite{totalrecall2015,totalrecall2016}. Text from the subject line and body were concatenated. As with RCV1-v2, documents with more than 512 WordPiece tokens were truncated, similar to the preprocessing steps used in prior works in email classification~\cite{shu2020learning}. Since the most recent replies and content are presented at the beginning of the email and the trailing parts are often duplicated from other emails, including only the leading passages are usually sufficient. 
A 50\% random sample of the remainder (137,062 documents) was used.
All 44 topics are used in the experiment. For consistency, we refer to these topics as categories in the later sections.

The RCV1-v2 news articles are professionally written texts with topics and vocabulary well covered by the book and encyclopedic text used to train BERT.  We view HRR on it as an in-domain task for BERT. The Jeb Bush emails (particularly from constituents) vary wildly in style and formality from message to message, and reference many Florida personalities, places, and issues likely to be poorly covered in the BERT pre-training materials. We therefore view it as an out-of-domain task for BERT.

\subsection{Software and Evaluation}

We implemented the active learning workflow with \texttt{libact}~\cite{libact}, an open-source active learning framework. For each category, a randomly selected positive example formed the sample for the first iteration.  On each subsequent iteration, 200 documents were sampled using active learning (either relevance feedback or least-confidence uncertainty sampling) by following experiment settings from prior HRR studies~\cite{yang2019icail, cost-structure-paper}. 

For BERT runs we used the \texttt{BERT-base-cased} model.\footnote{\url{https://huggingface.co/bert-base-cased}} 
Masked language model fine-tuning was done with the HuggingFace script \texttt{run\_mlm.py},\footnote{\url{https://github.com/huggingface/transformers/blob/master/examples/pytorch/language-modeling/run_mlm.py}} which uses ADAM with no weight decay and warm up period as the optimizer, and a learning rate of $5\times 10^{-5}$. To test the importance of language model fine-tuning, we vary it from
no language model fine-tuning to ten iterations over the corpus.

Then on each active learning iteration, we do classification fine-tuning using the ADAM optimizer with a linear weight decay of 0.01 with 50 warm up steps and initial learning rate of 0.001. All reviewed documents (including the ones previously reviewed) are used to fine-tune the model at each active learning iterations with 20 epochs. 
All hyperparameters were selected based on a pilot study on one selected category of each collection for maximizing the average R-Precision after 20 active learning iterations. The authors also experimented with fine-tuning the model with only the newly queried documents at each iteration, but the results were worse than fine-tuning on all labeled documents by a large margin. 

Logistic regression is served as the baseline in our study and is implemented with \texttt{scikit-learn}~\cite{scikit-learn} for comparison. It is widely used in HRR research and commercial software~\cite{yang2019sigir, brown2015peeking, bannach2019machine, cost-structure-paper}. We use the scikit-learn tokenizer and BM25 within document saturated term frequencies as feature values~\cite{robertson2009probabilistic, yang2019sigir}.  We use L2 regularization on the logistic losses, with penalty weight 1.0 and fit to convergence with default settings from scikit-learn.

For comparison with prior work, we report R-Precision, which is a metric that often reports in high recall retrieval studies~\cite{totalrecall2015, totalrecall2016, yang2019sigir}. Despite being an effectiveness measure that jointly considers precision and recall, it does not reflect the actual objective of the retrieval task, which is the reviewing cost. 

Therefore, our primary evaluation measure is the total optimal reviewing cost of the TAR run~\cite{cost-structure-paper}, which is the sum of reviewing the training documents and the documents ranked by the current classification model to fulfill the recall targe.  
The latter is referred to as the optimal amount of the second phase review and can be considered as an optimal penalty for the one-phase workflow~\cite{qbcb-paper, cost-structure-paper}. We report the minimal total cost that occurs during the 20 active learning iterations. 
Without loss of generality, we use 80\% recall target as an example, which is a widely used target in eDiscovery study. Higher targets such as 95\% yield similar results. 

To emphasize the importance of the underlying classification model in the iterative process, we evaluate with both the uniform cost structure (i.e., no reviewing cost difference between documents) and expensive training cost structure. Without loss of generality, we assume the training documents cost ten times more than documents reviewed during the mass reviewing phase as an example~\cite{cost-structure-paper}. The expensive training cost structure favors classification models that require less training data for optimizing the total cost, enabling us to distinguish the effectiveness of the classification model further.

\subsection{Hardware}

The active learning experiments are conducted on a cluster of 48 NVIDIA Titan RTX GPUs with 24 GB memory on each. One active learning run (one topic, one sampling strategy, one pretrained BERT model) took on average 18 hours. The entire set of experiments ($(45+44)\times5\times2=890$ runs) took around two weeks on our research cluster. 
The baseline experiments ran on a single CPU. All logistic regression runs ($(45+44)\times2=178$) took around one hour. A detailed running time analysis is presented in the next section. 
\section{Results and Analysis}

In this section, we aim to answer the following research questions: 
does language model fine-tuning improves the retrieval effectiveness? 
If so, what is the right amount? 
How much overhead are we paying for applying BERT? 

\begin{table}[t]

\newcommand{\minitab}[2][l]{\begin{tabular}{@{}#1@{}}#2\end{tabular}}

\centering
\caption{Averaged evaluation results on the in-domain RCV1-v2 collection and off-domain Jeb Bush collection over categories. Numbers in parentheses are the relative difference between the baseline logistic regression model (LR). Both uniform and expensive training cost (Exp. Train.) values are the relative cost difference between the BERT and the logistic regression models. Values larger than 1.0 indicate higher costs than the baseline. * indicates the statistical significance with 95\% confidence between the corresponding pretrained BERT model and the baseline conducted by paired t-test with Bonferroni corrections within each evaluation metric.}\label{tab:agg-metric}
\vspace{1em}
\setlength\tabcolsep{0.45em}
\begin{tabular}{lc|rr|rr|rr}
\toprule
         &  LMFT      & \multicolumn{2}{c|}{R-Precision ($\uparrow$)} & 
                        \multicolumn{2}{c|}{Uni. Cost ($\downarrow$)} & 
                        \multicolumn{2}{c}{Exp. Train.  ($\downarrow$)} \\
Collection & Epoch    &   Relevance &     Uncertainty & 
                          Rel. &     Unc. &
                          Rel. &     Unc. \\
\midrule

\multirow{6}{*}{\minitab[l]{In-domain\\RCV1-v2}} 
& LR &   0.788 (1.00) &   0.760 (1.00) &   1.000 &   1.000 &   1.000 &   1.000 \\
& 0  &   0.752 (0.95) &   0.756 (0.99) &   1.309 &   1.015 &   1.178 &  *0.873 \\
& 1  &   0.757 (0.96) &   0.768 (1.01) &   1.199 &   1.039 &   1.012 &   0.894 \\
& 2  &   0.759 (0.96) &   0.766 (1.01) &   1.289 &   1.028 &   1.067 &   0.890 \\
& 5  &   0.756 (0.96) &   0.784 (1.03) &   1.173 &   0.893 &   0.980 &  *0.844 \\
& 10 &   0.764 (0.97) &   0.765 (1.01) &   1.192 &   0.950 &   1.051 &  *0.878 \\
\midrule
\multirow{6}{*}{\minitab[l]{Off-domain\\Jeb Bush}} 
& LR &   0.904 (1.00) &   0.857 (1.00) &   1.000 &   1.000 &   1.000 &   1.000 \\
& 0  &  *0.724 (0.80) &  *0.719 (0.84) &   6.877 &   5.834 &  *2.717 &  *2.194 \\
& 1  &   0.811 (0.90) &   0.816 (0.95) &   4.678 &   2.896 &   1.756 &   1.413 \\
& 2  &   0.812 (0.90) &   0.808 (0.94) &   3.257 &   3.141 &   1.675 &   1.446 \\
& 5  &  *0.810 (0.90) &   0.813 (0.95) &   3.261 &   2.665 &   1.583 &   1.322 \\
& 10 &   0.805 (0.89) &   0.815 (0.95) &   3.922 &   2.943 &   1.601 &   1.361 \\

\bottomrule
\end{tabular}

\end{table}

\begin{table}[t]

\caption{Cost of RCV1-v2 categories in each bin under the expensive training cost structure. Values are the relative cost difference between the corresponding BERT and baseline models averaged over the five categories in each bin.}
\label{tab:rcv1-cost-breakdown}
\vspace{1em}
\setlength\tabcolsep{0.3em}
\centering
\resizebox{\textwidth}{!}{%
\begin{tabular}{ll|ccccc|ccccc}
\toprule
       &        & \multicolumn{5}{c}{Relevance} & \multicolumn{5}{|c}{Uncertainty} \\
Difficulty & Prevalence   
                &      0  &      1  &      2  &      5  &      10 &          0  &      1  &      2  &      5  &      10 \\
\midrule
Hard   & Rare   &     0.918 &  0.988 &  1.011 &  0.997 &  1.048 &       1.044 &  0.843 &  0.801 &  0.664 &  0.870 \\
       & Medium &     0.774 &  0.773 &  0.699 &  0.622 &  0.639 &       0.594 &  0.670 &  0.602 &  0.612 &  0.613 \\
       & Common &     0.832 &  0.856 &  0.850 &  0.798 &  0.755 &       0.815 &  0.849 &  0.842 &  0.755 &  0.751 \\
\midrule
Medium & Rare   &     0.932 &  0.916 &  0.904 &  0.784 &  0.951 &       0.770 &  0.903 &  0.868 &  0.794 &  0.828 \\
       & Medium &     1.275 &  1.311 &  1.293 &  1.175 &  1.229 &       1.065 &  1.199 &  1.203 &  1.088 &  1.211 \\
       & Common &     0.951 &  0.778 &  0.830 &  0.743 &  0.820 &       0.946 &  0.945 &  0.933 &  0.845 &  0.915 \\
\midrule
Easy   & Rare   &     1.688 &  1.225 &  1.362 &  1.430 &  1.540 &       0.587 &  0.638 &  0.702 &  0.632 &  0.621 \\
       & Medium &     1.897 &  1.189 &  1.182 &  1.103 &  1.263 &       1.073 &  0.936 &  1.015 &  1.069 &  0.982 \\
       & Common &     1.336 &  1.070 &  1.474 &  1.165 &  1.218 &       0.960 &  1.061 &  1.047 &  1.136 &  1.112 \\

\bottomrule
\end{tabular}
}
\vspace{-1.5em}

\end{table}

\subsection{Language Model Fine-Tuning}

Based on our experimental results, BERT with language model (LM) fine-tuning improves the effectiveness only when the domain of the collection aligns with the domain of the pretraining corpora. 
In Table~\ref{tab:agg-metric}, the reported cost is the average of the proportional relative cost differences between the baseline logistic regression results and the pretrained BERT model. Since the cost varies between categories, averaging the relative differences prevent the naturally harder tasks (with higher baseline cost) from diluting the aggregated values. The paired t-tests are still conducted between the raw cost with a null hypothesis of identical cost between the BERT and the baseline model. 
In RCV1-v2, BERT models provide roughly the same R-Precision (0.75 to 0.77) as the baseline logistic regression model regardless of the length of LM fine-tuning, suggesting similar quality at the top of the rank list. On the other hand, BERT models reduce the cost, especially when the training documents cost more to review, compared to the baseline model when the amount of LM fine-tuning is \textit{just right} (10\% to 15\% on average with expensive training cost structure). In our experiments, the \textit{goldilock} amount is five epochs. However, this amount varies with collection size and other characteristics of the task, which is discussed later in the section. 
Since reducing the total cost of TAR requires improving the overall rank list~\cite{cost-structure-paper}, these results suggest that the BERT model with five epochs of LM fine-tuning provides a consistent improvement on the entire ranking. 

If the target collection is off-domain compared to the original pre-trained corpora, BERT models cannot provide an effective classifier, even worse than simple linear logistic regression. The averaged values in the Jeb Bush collection suggest worse effectiveness (lower R-Precision and higher cost) despite that the differences are not statistically significant. However, the time overhead and computational burden of applying neural models such as BERT are massive compared to linear models. The inability to provide more effective retrieval results is already a failure. Note that the effectiveness of the BERT models could eventually improve over the baseline with more LM fine-tuning despite the decrement from five to ten epochs; the computational cost would be uneconomical. Running time analysis is presented later in this section. 

Therefore, applying BERT models to TAR is not guaranteed to lead to more effective retrieval results. The alignment of the domain between the collections and the amount of LM fine-tuning constitutes a considerable variation of the effectiveness, which is counterintuitive to the common wisdom that continuing fine-tuning would result in better results~\cite{gururangan2020don}. If \textit{just-right} hyperparameter is not available for the task, which is usually the case for real-world applications, applying BERT models could result in inferior results. 

\subsection{\textit{Just-Right} Varies Across Tasks}

The 45 categories selected from RCV1-v2 enable further analysis into the effect of the task characteristics. Table~\ref{tab:rcv1-cost-breakdown} demonstrates the averaged relative cost differences compared to the baseline model in each category bin under the expensive training cost structure. Since each bin only contains five runs (five categories), statistical tests are %
non-indicative; hence omitted.

For relevance feedback where training documents are selected from the top of the rank, BERT models usually perform similarly to logistic regression models with a few exceptions. BERT models are more helpful in hard categories than easy ones since the relevancy is often beyond simple token matching in the hard ones, yielding a 20\% to 30\% cost reduction. However, when the task is hard and the relevant documents are rare, BERT models are no better than simple linear models, even with more LM fine-tuning.

For uncertainty sampling, where the training documents are ones that the model is the least certain about (with predicted probability around 0.5), BERT models provide a substantial improvement of 20\% to 40\% cost reduction in both hard and rare categories. These results indicate that BERT models are still more effective in challenging situations -- either extremely unbalanced training set or relevancy requires subtle semantic understanding. These are cases where linear models tend to fail if no specific treatments to the collection are made.

However, even in these cases where BERT models demonstrate a clear advantage over the linear models, the amount of LM fine-tuning is still critical. The optimal length of LM fine-tuning varies across difficulty and prevalence bins, which were developed by \citet{cost-structure-paper}. For example, the best performing pre-trained model for the \textit{hard-medium} bin is no LM fine-tuning (0.5935, i.e., 41\% cost reduction). However, LM fine-tuning for five epochs gives us the lowest cost (0.6637) and seems to be the minimum for \textit{hard-rare}. For \textit{hard-common}, more fine-tuning tends to be consistently improving the model with the lowest cost (0.7512) occurred at ten epochs in our experiment. The trend is different for medium and easy difficulty bins. 

\begin{figure}[]
    \centering
    \includegraphics[width=\linewidth]{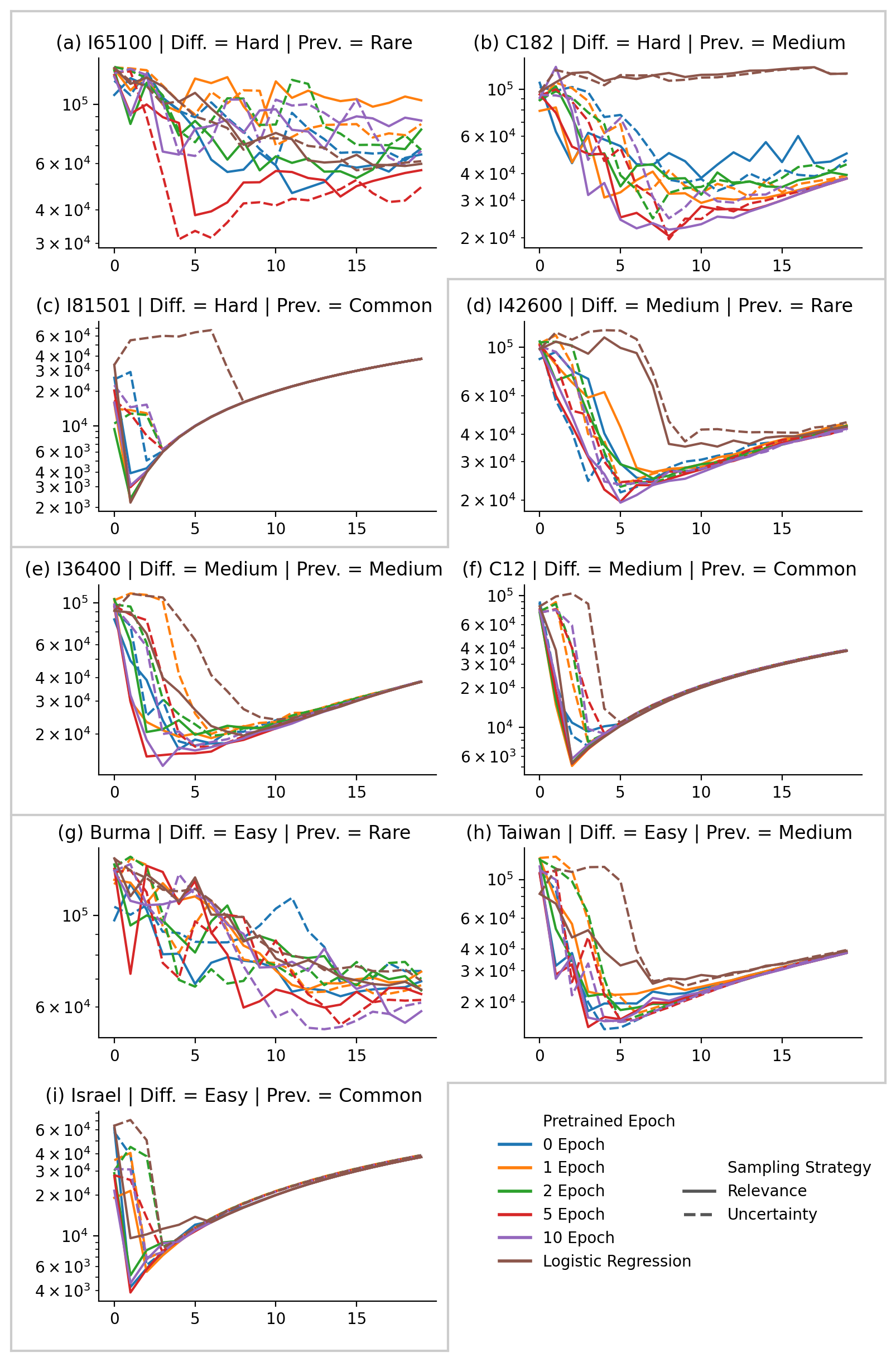}
    \caption{Example total cost of TAR runs on RCV1-v2 collection over the rounds with expensive training cost structure. The y-axis is the total cost in log-scaled to demonstrate the differences and the x-axis is the number of TAR rounds. }
    \label{fig:rcv1-cost-breakdown}
\end{figure}
\begin{table}[t]

\newcommand{\minitab}[2][l]{\begin{tabular}{@{}#1@{}}#2\end{tabular}}
\newcommand{\z}{\phantom{0}}

\caption{Running time in minutes. Running time for LM fine-tuning (LMFT) is agnostic to the categories. Time reported for TAR is the average running time for each category to complete a 20-iteration TAR process, which consists of 20 classification fine-tuning (or training for logistic regression) and scoring the entire collection. Values in parentheses are the standard deviations of the averaged time.  }
\label{tab:running-time}
\vspace{1em}
\setlength\tabcolsep{0.55em}
\centering
\begin{tabular}{lc|rrr|rrr}
\toprule
           & LMFT & \multicolumn{3}{c}{Relevance} 
                      & \multicolumn{3}{|c}{Uncertainty} \\
Collection & Epoch  &  LMFT &  TAR &  Total
                    &  LMFT &  TAR &  Total \\
\midrule

\multirow{6}{*}{\minitab[l]{In-domain\\RCV1-v2}} 
& 0  &      -- &  1095 &  1095 (31.49) &     -- &  1098 &  1098 (28.97) \\
& 1  &      98 &  1094 &  1192 (17.55) &     98 &  1102 &  1200 (20.57) \\
& 2  &     196 &  1096 &  1292 (20.53) &    196 &  1100 &  1296 (28.33) \\
& 5  &     490 &  1103 &  1593 (23.57) &    490 &  1103 &  1593 (19.93) \\
& 10 &     980 &  1101 &  2081 (20.26) &    980 &  1105 &  2085 (20.90) \\ 
& LR &      -- &  0.32 & 0.34 (\z0.04) &     -- &  0.38 & 0.38 (\z0.05) \\
\midrule
\multirow{6}{*}{\minitab[l]{Off-domain\\Jeb Bush}} 
& 0  &      -- &   999 &   999 (19.02) &     -- &  1008 &  1008 (19.48) \\
& 1  &      98 &  1002 &  1100 (16.56) &     98 &  1003 &  1101 (24.78) \\
& 2  &     196 &  1002 &  1198 (15.03) &    196 &  1002 &  1198 (21.80) \\
& 5  &     490 &  1007 &  1497 (19.34) &    490 &  1004 &  1494 (27.37) \\
& 10 &     981 &   996 &  1977 (22.48) &    981 &  1006 &  1987 (26.67) \\
& LR &      -- &  0.33 & 0.33 (\z0.04) &     -- &  0.41 & 0.41 (\z0.06) \\

\bottomrule
\end{tabular}

\vspace{-1.5em}

\end{table}

Beyond minimum cost during the run, the trajectory of cost over the iterations also varies among different numbers of LM fine-tuning epochs. 
For the \textit{hard-rare} category (\texttt{I65100}) in Figure~\ref{fig:rcv1-cost-breakdown}(a), the transition from the trajectory of 1 epoch of LM fine-tuning to 2 is not smooth and the shape is nowhere similar. The \textit{hard-common} category (\texttt{I81501} in Figure~\ref{fig:rcv1-cost-breakdown}(c)) also convey no clear relationship between different number of LM fine-tuning epochs. 

While BERT models provide significant improvement over the failure cases such as the \textit{medium-rare} category (\texttt{I42600}, Figure~\ref{fig:rcv1-cost-breakdown}(d)) and \textit{hard-medium} category (\texttt{C182}, Figure~\ref{fig:rcv1-cost-breakdown}(b)), the trajectory is nearly identical for the easy categories regardless of the LM fine-tuning epochs, especially with relevance feedback.

Despite making no clear conclusion on the optimal amount of LM fine-tuning, we observe that this hyperparameter is critical and independent of the collection. All TAR runs in Table~\ref{tab:rcv1-cost-breakdown} are based on the same 20\% subset of RCV1-v2 collection but with different categories. This poses a challenge for TAR practitioners when applying BERT or potentially other transformer-based classification models to projects: the joint effect of this hyperparameter and the characteristics of the task is so large that it ranges from extremely helpful (50\% cost reduction in \textit{hard-medium} categories using uncertainty sampling without LM fine-tuning) to large cost overhead (89\% cost overhead in \textit{easy-medium} categories using relevance feedback without LM fine-tuning). Understanding the characteristics of the task remains crucial but challenging without sufficient annotations, which is one of the purposes for applying TAR. 

\subsection{Running Time}

Finally, we analyze the running time of the TAR runs. In Table~\ref{tab:running-time}, the computational overhead of applying BERT is massive. While the training and scoring of the collection during TAR using logistic regression takes on average 20 to 25 seconds (0.32 to 0.41 minutes), the BERT model takes around 18 hours (1100 minutes). The majority of the time was spent on scoring the collection, which takes around 40 minutes at each iteration. The LM fine-tuning is done before the TAR iterative process, taking around 100 minutes per epoch for the collections we experimented with. 

In real high recall retrieval applications where the iterative process spans weeks or even months, each round of reviewing documents takes around half a day. Adding one hour overhead to each iteration is potentially acceptable. However, for smaller projects, this significant time overhead could directly prevent BERT from applying. The computational cost for applying BERT is also not amortized to millions of queries after deployment. Spending 18 hours training a single-usage model in exchange for a mild effectiveness improvement could be unnecessary overhead for many HRR projects. 

\section{Summary and Future Works}

We evaluated the effectiveness of TAR with pre-trained BERT as the underlying predictive model. Before entering active learning, the pre-trained BERT model is fine-tuned by the masked language modeling objective with several epochs. Through experiments, we show that the amount of LM fine-tuning is critical even on an in-domain task. For tasks with out-of-domain text, as compared to the BERT model pre-training corpora, LM fine-tuning requires more training, potentially with other similar corpora. Without proper LM fine-tuning, BERT models underperform typical linear models used with TAR. However, our experiments also show that category characteristics also impact how beneficial the BERT models are and the large computational overhead might discourage the application of BERT in real-world HRR projects. 

As the first study of applying transformer models to TAR, there is still much to explore in this area. %
In the future, we will investigate a wider variety of HRR tasks and sampling strategies that are designed for neural models such as Monte Carlo Dropout~\cite{gal2016dropout} and Discriminative Active Learning~\cite{gissin2019discriminative}. A comprehensive approach for handling documents with more than 512 tokens should also be studied.  Pre-training a transformer model with large email corpora would benefit the community as many eDiscovery tasks are working on emails.  Whether the pre-training corpora would carry biases into the final retrieval results in each TAR project is also demanding for future research.

\bibliographystyle{splncs04nat}
\bibliography{ms}

\end{document}